# A database of travel-related behaviors and attitudes before, during, and after COVID-19 in the United States

Rishabh Singh Chauhan[1], Matthew Wigginton Conway[2], Denise Capasso da Silva[3], Deborah Salon[2], Ali Shamshiripour[1], Ehsan Rahimi[1], Sara Khoeini[3], Abolfazl (Kouros) Mohammadian[1], Sybil Derrible[1], and Ram Pendyala[3]

1. Department of Civil, Materials, and Environmental Engineering, University of Illinois at Chicago, IL, USA;
2. School of Geographical Sciences and Urban Planning, Arizona State University, Tempe, AZ, USA;
3. School of Sustainable Engineering and the Built Environment, Arizona State University, Tempe, AZ, USA

## Abstract

The COVID-19 pandemic has impacted billions of people around the world. To capture some of these impacts in the United States, we are conducting a nationwide longitudinal survey collecting information about activity and travel-related behaviors and attitudes before, during, and after the COVID-19 pandemic. The survey questions cover a wide range of topics including commuting, daily travel, air travel, working from home, online learning, shopping, and risk perception, along with attitudinal, socioeconomic, and demographic information. The survey is deployed over multiple waves to the same respondents to monitor how behaviors and attitudes evolve over time. Version 1.0 of the survey contains 8,723 Wave 1 responses that are publicly available. This article details the methodology adopted for the collection, cleaning, and processing of the data. In addition, the data are weighted to be representative of national and regional demographics. This survey dataset can aid researchers, policymakers, businesses, and government agencies in understanding both the extent of behavioral shifts and the likelihood that changes in behaviors will persist after COVID-19.

## Background & Summary

The COVID-19 pandemic has spread across the world, infecting tens of millions and killing over one million people[1]. By March 2021, the United States (U.S.) had recorded the highest number of confirmed COVID-19 cases and COVID-19 related deaths in the world[1]. Since social distancing is one of the most effective measures in containing the spread of the infection[2], several U.S. states issued various restrictions including stay at home orders. Moreover, numerous restaurants and bars closed for dine-in services, various recreation facilities were shut down, many offices and schools switched from meeting in-person to meeting online, and travel restrictions were imposed. These measures had a profound impact on how people in the U.S. went about their daily lives.

To understand the current and future impacts of the pandemic, we conducted a nationwide online survey. The goal of the survey is to capture attitudes and shifts in travel-related choices of people across the nation both during the pandemic and once COVID-19 is no longer a threat. The data are shared publicly in order to help government agencies and businesses prepare for the future. We are conducting additional survey waves with the same respondents to monitor how people's choices evolve over the course of the pandemic and beyond.

An early version of the survey took place from April to June 2020, when the stay at home orders were in place in most parts of the country[3,4]; this portion of the data collection is referenced as *Wave 1A*. A slightly-modified larger-scale survey, *Wave 1B*, was deployed between late June and October 2020. Subsequent survey waves are being conducted as the situation evolves. The collected data are released as they become available and necessary procedures for cleaning, documenting, and weighting the data are completed. This procedures for data processing are detailed in this paper. The present article focuses on data from the first wave of the survey.

In the months following the beginning of the spread of COVID-19, several efforts have been made to collect data related to COVID-19. In fact, many datasets have been compiled, specifically on COVID-19 testing[5], medical imaging of COVID-19 cases[6], the timeline of government interventions[7], policy announcements[8], implementation and relaxation of public health and social measures[9], epidemiological data[10], mobility-related data[11], and out-of-home activity information[12], to name a few. Researchers also turned to social media platforms, like Twitter and Instagram, to gather COVID-19-related data[13–16]. Furthermore, several surveys have been conducted to measure the impacts of the pandemic[17–19], some of which are now released for public use[20,21].

Our survey data are different from most others in several ways. First, it is comprehensive insofar as it includes data about a wide range of topics including commuting, daily travel, air travel, working from home, online learning, shopping, attitudes, risk perception, and socioeconomic and demographic details. Second, it captures detailed information about behaviors before and during the COVID-19 pandemic, as well as the choices that people expect to make when the COVID-19 virus is no longer a threat. Third, it was collected from respondents across the U.S., covering diverse socio-economic backgrounds, professions, education levels, and ages. Fourth, the survey is a true longitudinal panel survey, collecting data in multiple waves from the same individuals at regular intervals. Finally, the data are made publicly available to promote data-driven analysis and research.

The next section describes the data collection methodology, the questions included in the survey, the survey deployment process, and the participant recruitment strategy. Next, the data records section describes the data file types and metadata. Subsequently, the technical validation section explains the procedure for the survey data cleaning and weighting. Lastly, the final section provides additional notes for data users.

## Methods

### Ethical Compliance

Our study protocol was approved by both Arizona State University (ASU) and University of Illinois at Chicago (UIC) Institutional Review Board offices. Participants were informed that their participation is voluntary, and that their responses are shared anonymously. An online informed consent was obtained from everyone who responded to the survey.

### Survey Questions

The data were collected through an extensive online survey with over 120 questions. The survey questions can be broadly divided into three categories: (1) retrospective questions focusing on the period before COVID-19, (2) questions about the period during COVID-19, and (3) prospective questions on respondent expectations for a future period in which COVID-19 is no longer a threat. The questions cover a wide variety of subjects including commuting habits, discretionary travel choices, work-related questions, study-related questions, shopping, dining, and so on – all before, during, and expected after the pandemic.

The survey questions can be classified into eight categories based on question subject type, namely: demographics, work, study, shopping and dining, transportation, and general attitudes. Table 1 describes each of these categories.

Table 1: Survey Sections Description

| Survey Sections | Description |
|---|---|
| Demographics | age; gender; race; ethnicity; income; household information; access to high-speed internet/mobile phone; and, physical disability. |
| Transportation | available vehicles in the household; vehicles purchased or sold after the beginning of COVID-19; mode usage frequency before COVID-19, during COVID-19, and expectation for the post-COVID-19 period; transportation to school before and after COVID-19; commute to work before and during COVID-19; and air travel before COVID-19 and expectations for the post-COVID-19 period. |
| Work | work from home options and work productivity – before and during COVID-19; and work from home expectation for the post-COVID-19 period. |
| Study | learning affected during COVID-19. |
| Shopping and Dining Behavior | shopping and dining preferences pre-COVID-19, during COVID-19, and expectations for the post-COVID-19 period. |
| Pandemic experience | new ways of living to keep after the pandemic; negative impacts from the pandemic; and COVID-19 testing status. |
| Attitudes | risk perception; socializing preferences; attitudes towards the environment; residential preferences; technology savviness; and views on COVID-19. |
| Social resources | social resources to which respondents can turn to for things such as finding a job or borrowing money. |

**Survey Recruitment**

From April to mid-June 2020, initial Wave 1A responses were collected from a convenience sample via mailing lists, social media outreach, and mainstream media articles. A total of 1,110 responses were collected during this phase.

From late June onward, Wave 1B, the modified version of the survey, was deployed through survey invitations sent to a random email list purchased from a data marketing company. The list contained 350,000 email addresses belonging to people in 24 metropolitan areas across the U.S., as well as the state of Ohio (see Figure 1). We purchased 100,000 additional email addresses of people randomly selected from across the country, including rural areas and excluding the areas covered by the first 350,000 emails. A total of 1,116 responses were received from the email list. Unfortunately, major email service providers quickly began marking our survey invitations as spam, while some smaller providers did not. While we took several steps to mitigate this issue, including changing the wording of the emails, changing the source of the emails (a uic.edu, asu.edu, or covidfuture.org email address), we were ultimately not able to fully solve this problem and saw a lower response rate from individuals with addresses from major email providers.

Survey invitation emails were also sent to an additional list of approximately 39,000 email addresses from the Phoenix metropolitan area purchased for a previous survey effort[22]. This list yielded 782 responses. The survey invitation emails were sent using Amazon Web Services (AWS) and through the Qualtrics platform. Every 20th respondent who was invited through the purchased email addresses received a $10 incentive as a gift card. Respondents also had the option to donate their survey incentive to a charity. Invitees received two reminders as part of efforts to maximize response rates.

An additional 5,250 responses to the Wave 1B survey were collected through a Qualtrics Online Panel. Qualtrics recruits these respondents from a variety of panels maintained by other firms and uses quota sampling to recruit respondents that are demographically representative of the nation. The Qualtrics quotas were set to collect information from 20 U.S. metropolitan areas, mostly consistent with the metropolitan areas sampled from the purchased email list, as well as the states of Ohio, Utah, North Carolina, upstate New York, and rural areas. In order to obtain samples that would represent the population in each of the selected geographies, quotas were imposed in the Qualtrics online panel subsample to guarantee representation based on income, age, race and ethnicity, and education. We requested all respondents to provide their email addresses in order to recontact them for subsequent survey waves. Since the Qualtrics respondents are professional survey takers, we designated most questions as mandatory, and we included attention check questions, which are shown to improve response quality[23].

The distribution of responses by geography, as well as the targeted metropolitan areas, are shown in Figure 1. Figure 2 shows the distribution of responses by recruitment method, available in the "org" variable in the dataset. The geographical targets were chosen based on geographic and metropolitan area size diversity, as well as the state of the virus spread in May 2020.

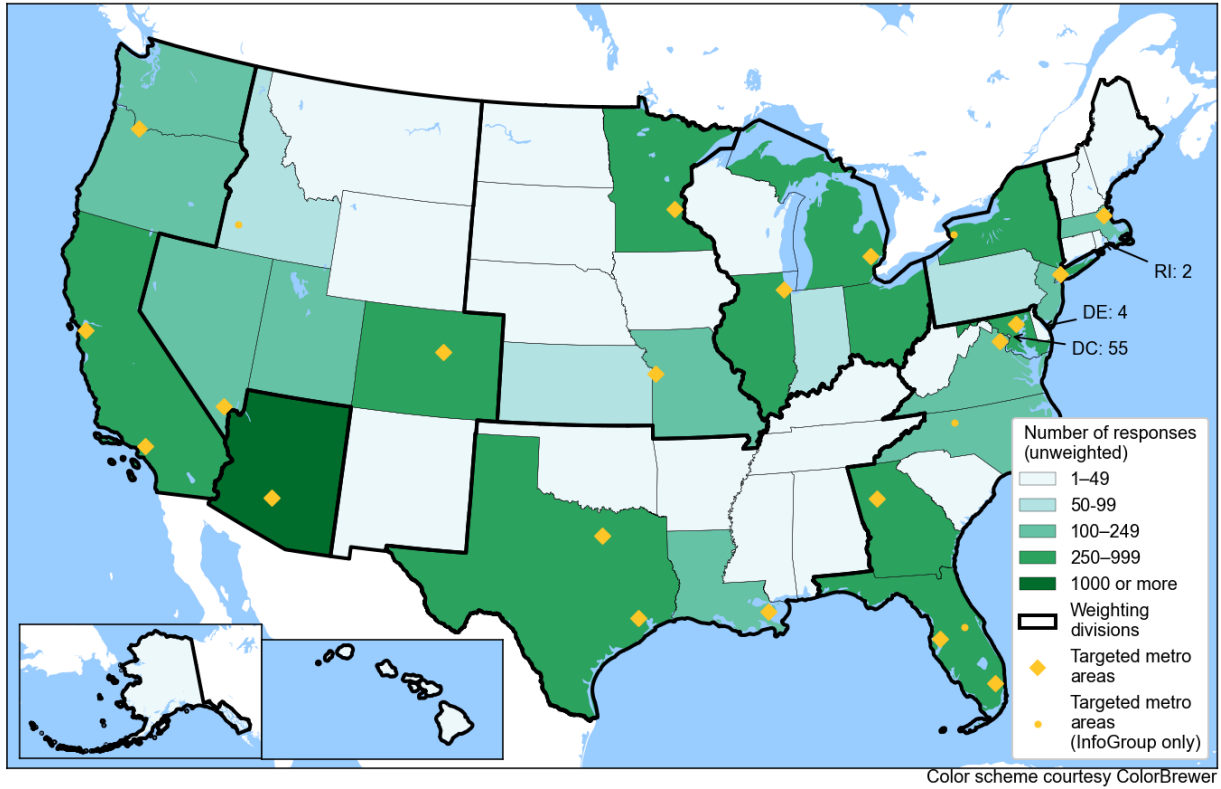

Figure 1: Distribution of survey respondents by the state of residence for survey dataset version 1.0. Alaska and Hawai'i are in the same weighting division as California, Oregon, and Washington.

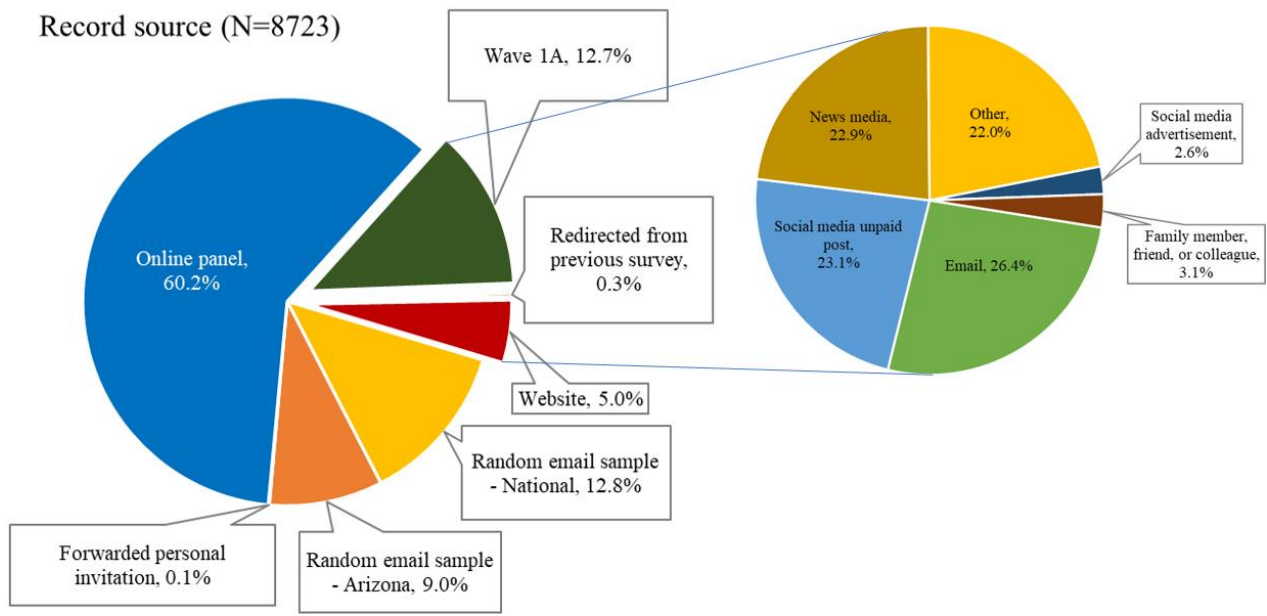

Figure 2: Distribution of Record by Source (from the survey dataset version 1.0).

Figure 1 shows the distribution of survey respondents across the U.S. (50 states and the District of Columbia). Following our recruitment strategy, a greater number of responses come from larger and more urban states. Arizona is overrepresented due to the oversample of Arizona respondents in the email-based deployment. The respondents from the initial Wave 1A sample are also more likely to hail from Arizona as the Arizona State University survey team's network is heavily Arizona-based. When the data are weighted, any geographic discrepancies at the census division level are controlled and overrepresentation of Arizona is controlled/corrected separately.

**Additional survey waves**

To monitor how people's attitudes and behaviors evolve, survey respondents are contacted again with at least two shorter follow-up surveys, approximately four months apart. In addition, as of the date of publication of this manuscript, we are continuing to collect responses from Wave 1B from new respondents who visit the project website.

## Data Records

The survey dataset[24] can be accessed from the ASU Dataverse at: https://doi.org/10.48349/ASU/QO7BTC. The dataset is available in CSV (comma-separated value) format. Since the data will be updated periodically, the data are versioned—in this article, results from the survey dataset version 1.0 are reported. The dataverse also contains the database codebook containing the metadata and explaining the variables. The codebook contains a changelog for each new version.

The respondents to Waves 1A and 1B received similar but not identical surveys. We have merged the responses to these two versions of the survey into the final dataset wherever possible. For some variables, the questions were identical, whereas for other variables, harmonization of similar responses was required. In the dataset, variables ending in '_harm' are harmonized between the two datasets, variables ending in '_w1a' are available only for Wave 1A respondents, variables ending in '_w1b' are available only for respondents from our Qualtrics Online Panel, purchased email lists, and anyone who found the survey via the COVIDFuture web site or email lists after June 19, 2020 (start date of Wave 1B). Variables with no suffix were asked the same way between the two surveys, and no harmonization was necessary. We also provide a file containing only Wave 1B responses and variables, which simplifies analysis of the Wave 1B data.

## Technical Validation

**Data Cleaning**

To monitor respondents' attention to survey questions in the Qualtrics online panel, attention check questions were included. Respondents were allowed to miss one attention check and be given an opportunity to answer that section again. If they missed an attention check twice, or both attention checks once, their survey was terminated.

We additionally undertook several quality checks to help ensure that the collected data were valid. We removed any respondents who reported that they shop for groceries both in-store and online every day, or expect to after the pandemic, as these are likely to be invalid responses.

We also removed respondents who reported strongly agreeing or strongly disagreeing with all COVID-related attitudes, as some of these were worded positively and some negatively. Several additional quality checks were undertaken in the Qualtrics Online Panel as part of Qualtrics' data cleaning process, including a check for people finishing the survey too quickly.

Respondents that did not report a state of residence, reported living outside the 50 states and the District of Columbia, or did not provide answers to all of the control variables used in the data weighting process described in the next section were removed from the data. Due to this restriction, 558 records with missing control variable information, 59 records with missing home location, and one response from Puerto Rico were not included in the final dataset encompassing responses received through October 14, 2020. Further steps in data preparation will include imputation of missing data, which will allow for some of these omitted records to be recovered in the next version of the dataset. Among the respondents who were not included in the dataset due to missing control variable information, there are 34 respondents who declared their gender as Other; these respondents could not be included because the Census offers no control marginals to weight these records. Further data weighting procedures will attempt to incorporate non-binary gendered individuals on the dataset. Due to the data cleaning and filtering process applied to responses obtained through October 14, 2020, a total of 618 records were not included in the published dataset.

**Data Weighting**

Because the raw data are not fully representative of the U.S. population, weights were calculated using the following control variables: age, education, gender, Hispanic status, household income, presence of children, and number of household vehicles. The weighting procedure accounts for the true population characteristics at the person level. Household-level variables (i.e., income, presence of children, and number of vehicles) were controlled at the person level as well. For example, the marginal distribution used for presence of children refers to the share of adults aged 18 years and older living in household with children, instead of the share of households that have children as it is usually represented. Those marginal distributions were computed using data from the Integrated Public Use Microdata Sample and the American Community Survey (ACS) 2018 1-year data[25] using the sample 18 and older in each of the weighting region boundaries. A noteworthy consequence of this approach is that adjusted household weights are necessary to evaluate household-level characteristics since individuals from larger households are more likely to be represented in the survey (given there are more individuals in these households), and thus have a higher probability of being selected. Weights for household-level analysis can be computed by dividing the person-level weight (provided in the data) by the number of adults in the household.

The national sample was divided into nine regions based on the reported home state (Table 2). Each region's sample was then weighted to match the distributions observed in ACS 2018 1-year estimates[25], meaning that the survey is demographically representative at the level of each region as well as the entire U.S. The unweighted and weighted survey results are shown in Table 3; the weighted results closely replicate population distributions, with inevitable minor deviations on variables that were not controlled in the weighting process.

Weights were calculated using iterative proportional fitting (IPF) procedures embedded within the synthetic population generator PopGen2.0[26–28]. Univariate marginal control distributions were derived from the Integrated Public Use Microdata Sample, American Community Survey (ACS) 2018 1-year data[25].

Table 2: Geographic Resolution of Weighting Procedure

| Census Division | Description |
|---|---|
| Division 1 | New England (Connecticut, Maine, Massachusetts, New Hampshire, Rhode Island, Vermont) |
| Division 2 | Middle Atlantic (New Jersey, New York, Pennsylvania) |
| Division 3 | East North Central (Indiana, Illinois, Michigan, Ohio, Wisconsin) |
| Division 4 | West North Central (Iowa, Nebraska, Kansas, North Dakota, Minnesota, South Dakota, Missouri) |
| Division 5 | South Atlantic (Delaware, District of Columbia, Florida, Georgia, Maryland, North Carolina, South Carolina, Virginia, West Virginia) |
| Divisions 6-7 | West and East South Central (Alabama, Kentucky, Mississippi, Tennessee, Arkansas, Louisiana, Oklahoma, Texas) |
| Division 8, modified | Mountain, except Arizona (Colorado, Idaho, New Mexico, Montana, Utah, Nevada, Wyoming) |
| Division 8, Arizona | Arizona state only |
| Division 9 | Pacific (Alaska, California, Hawaii, Oregon, Washington) |

Table 3: Comparison of unweighted and weighted distributions of key sociodemographic variables with true population at the National level (Waves 1A and 1B combined)

| | | Unweighted (N=8,723) | Weighted (N=8,723) | ACS 2018 1-year estimates |
|---|---|---|---|---|
| *Person Characteristics* | | | | |
| Age | 18-29 years | 16.4% | 21.2% | 21.2% |
| | 30-44 years | 26.4% | 25.1% | 25.1% |
| | 45-64 years | 24.4% | 24.9% | 24.9% |
| | 65 years and above | 32.8% | 28.8% | 28.8% |
| Gender | Male | 37.5% | 48.7% | 48.7% |
| | Female | 62.5% | 51.3% | 51.3% |
| Education | High School Degree or Less | 16.2% | 39.3% | 39.3% |
| | Some College or Associate's Degree | 29.2% | 30.6% | 30.6% |
| | Bachelor's Degree or Higher | 54.6% | 30.1% | 30.1% |
| Employment* | Employed | 62.6% | 60.6% | 61.5% |
| | Not employed | 37.4% | 39.4% | 38.5% |
| Ethnicity | Hispanic | 10.4% | 16.2% | 16.2% |
| | Not Hispanic | 89.6% | 83.8% | 83.8% |
| Race* | White | 81.3% | 78.6% | 73.8% |
| | Non-White | 18.7% | 21.4% | 26.2% |
| *Person Level Controls of Household Characteristics* | | | | |
| Size* | 1 | 18.7% | 10.7% | 16.5% |
| | 2 | 38.3% | 32.3% | 32.8% |
| | 3 | 17.7% | 21.1% | 19.0% |
| | 4 or larger | 25.3% | 35.8% | 31.7% |
| Presence of children | Present | 27.3% | 35.9% | 35.9% |
| | Not present | 72.7% | 64.1% | 64.1% |
| Tenure* | Homeowner | 63.8% | 64.1% | 65.4% |
| | Not homeowner | 36.2% | 35.9% | 34.6% |
| | 0 | 7.0% | 9.3% | 9.3% |

| Vehicles available | 1 | 37.2% | 22.8% | 22.8% |
| --- | --- | --- | --- | --- |
| | 2 | 38.3% | 37.5% | 37.5% |
| | 3 or more | 17.5% | 30.4% | 30.4% |
| 2019 income before taxes | Less than $35,000/year | 24.1% | 20.7% | 20.7% |
| | $35,000 to $99,999/year | 42.5% | 41.8% | 41.8% |
| | More than $100,000/year | 33.4% | 37.5% | 37.5% |

*Variable was not controlled in weight computation

## Usage notes

Since the survey will be followed by at least two follow-up survey waves, the database will be updated periodically after the data for each wave is collected, cleaned, and weighted. Each version of the data will be uploaded to the ASU Dataverse and assigned a new DOI number, and all previous versions will remain available to promote reproducibility.

The weights were developed to produce a sample that is representative of the U.S. population, as well as representative of nine divisions within the U.S.: eight census regions (with East and West South Central combined due to small samples in these regions), and a separate category for Arizona due to its large number of respondents. The weights are not guaranteed to produce a representative sample for other (smaller) geographies. When evaluating subsamples at a finer geography (e.g., state or metropolitan area), data users should compare marginal distributions of key demographic variables with the census, and re-weight the data if needed to be representative of the area being analyzed.

Some questions differ between Waves 1A and 1B. Therefore, we have weighted the dataset twice: once including all respondents (Waves 1A and 1B), and once excluding respondents to the Wave 1A sample. Data users should use the Wave 1B weights whenever using variables that are not present in the convenience sample. Since Wave 1A data deviates significantly in terms of population representativeness[4], there are no weights for questions asked only of Wave 1A respondents. In the file with only Wave 1B responses, only Wave 1B weights are presented.

This unique dataset provides insights on attitudes and behaviors not just before and during pandemic, but also on what might be expected after the pandemic. Possible use cases include modeling of during-pandemic and longer-term changes in mode use, air travel, transit ridership, work from home, and traffic congestion (especially for peak period traffic planning) Published uses of this dataset are documented in Capasso da Silva *et al.* [29], Chauhan *et al.* [30], Mirtich *et al.* [31], and Salon *et al.* [32].

## Code availability

No codes were developed for this research.

**Acknowledgements**

This research was supported in part by the National Science Foundation (NSF) RAPID program under grants no. 2030156 and 2029962 and by the Center for Teaching Old Models New Tricks (TOMNET), a University Transportation Center sponsored by the U.S. Department of Transportation through grant no. 69A3551747116, as well as by the Knowledge Exchange for Resilience at Arizona State University. This COVID-19 Working Group effort was also supported by the NSF-funded Social Science Extreme Events Research (SSEER) network and the CONVERGE facility at the Natural Hazards Center at the University of Colorado Boulder (NSF Award #1841338) and the NSF CAREER award under grant no. 155173. Any opinions, findings, and conclusions or recommendations expressed in this material are those of the authors and do not necessarily reflect the views of the funders.

**Author Contribution**

RP, AM, SD, DS, and SK planned the project. DS, MC, DCS, RC, ER, and AM prepared the survey questions. MC, DCS, and DS designed the survey flow logic. RC, DCS, MC, DS, and SD deployed the survey. MC and DCS performed data cleaning and survey data analysis. DCS weighted the dataset. MC and DS worked on sending out the incentives to the selected respondents. RC prepared the first draft. All the authors made significant contributions to manuscript editing and approving the final version of the manuscript.

**Competing Interests**